\documentclass[fleqn,usenatbib]{mnras}
\usepackage[utf8]{inputenc}

\usepackage{acronym}
\usepackage{ae,aecompl}
\usepackage{booktabs}
\usepackage{dcolumn}
\usepackage{graphicx}
\usepackage{amsmath,amsfonts,amssymb}
\usepackage{mathrsfs}
\usepackage[normalem]{ulem}

\usepackage[T1]{fontenc}

\def\gtaprx {\lower .14ex\hbox{\rlap{\raise .9ex\hbox{\hskip .3ex
	{\ifmmode{\scriptscriptstyle >}\else
		{$\scriptscriptstyle >$}\fi}}}
	\kern -.4ex{\ifmmode{\scriptscriptstyle \sim}\else
		{$\scriptscriptstyle\sim$}\fi}}}
\def\ltaprx {\lower .14ex\hbox{\rlap{\raise .9ex\hbox{\hskip .3ex
	{\ifmmode{\scriptscriptstyle <}\else
		{$\scriptscriptstyle <$}\fi}}}
	\kern -.4ex{\ifmmode{\scriptscriptstyle \sim}\else
		{$\scriptscriptstyle\sim$}\fi}}}
\newcommand{\s}{ \, {\rm s} }

\newcommand{\km}{{\, \rm km}}

\newcommand{\nue}{\nu_{\rm e}} 
\newcommand{\nueb}{{\bar \nu}_{\rm e}} 
\newcommand{\nux}{\nu_x} 
\newcommand{\nuxb}{\bar \nu_x}

\newcommand{\del}[2]%
{\frac{\mathrm{d}{#2}}{\mathrm{d}{#1}}}
\newcommand{\Del}[2]%
{\frac{D{#2}}{D{#1}}}\newcommand{\ddel}[2]%
{\frac{\mathrm{d}^2{#2}}{\mathrm{d}{#1}^2}}

\newcommand{\pdel}[2]%
{\frac{\partial{#2}}{\partial{#1}}}
\newcommand{\pddel}[2]%
{\frac{\partial^2{#2}}{\partial{#1}^2}}

\newcommand{\Ms}{M_{\odot}}

\let\oldhref\href
\renewcommand{\href}[2]{\oldhref{#1}{\hbox{#2}}}

\title[Nucleosynthesis with FFC in CCSN]
{Explosive nucleosynthesis with fast neutrino-flavor conversion in core-collapse supernovae}

\author[S. Fujimoto and H. Nagakura]{Shin-ichiro Fujimoto$^{1}$
\thanks{E-mail: fuji@kumamoto-nct.ac.jp},
Hiroki Nagakura$^{2}$
\\
$^{1}$National Institute of Technology, Kumamoto College, Kumamoto 861-1102, Japan
\\
$^{2}$Division of Science, National Astronomical Observatory of Japan, 2-21-1 Osawa, Mitaka, Tokyo 181-8588, Japan\\
}

\date{Accepted XXX. Received YYY; in original form ZZZ}

\pubyear{2022}

\begin{document}
\label{firstpage}
\pagerange{\pageref{firstpage}--\pageref{lastpage}}
\maketitle

%% Abstract of the paper
\begin{abstract}
Fast neutrino ($\nu$)-flavor conversion (FFC) is a possible game-changing ingredient in core-collapse supernova (CCSN) theory.
In this paper, we examine the impact of FFC on explosive nucleosynthesis by including the effects of FFC in conjunction with asymmetric $\nu$ emission into nucleosynthetic computations in a parametric way. 
We find that the ejecta compositions are not appreciably affected by FFC for elements lighter than Co
while the compositions are influenced by FFC for the heavier elements.
We also find that the role of FFC varies depending on the asymmetric degree of $\nu$ emission ($m_{\rm asy}$) and the degree of $\nu$-flavor mixing.
The impact of FFC is not monotonic to $m_{\rm asy}$;
The change in the ejecta composition increases for higher $m_{\rm asy}$ up to $\sim 10 \%$ compared with that without FFC,
whereas FFC has little effect on the nucleosynthesis in very large asymmetric $\nu$ emission ($\gtrsim 30 \%$). 
Our results suggest that FFC facilitates the production of neutron-rich ejecta in most cases, although it makes the ejecta more proton-rich if anti-$\nu$ conversion is more vigorous than that of $\nu$. 
The key ingredient accounting for this trend is $\nu$ absorption, whose effects on nucleosynthesis can be quantified by simple diagnostics.
\end{abstract}

%% Select between one and six entries from the list of approved keywords.
%% Don't make up new ones.
%% keyword1 -- keyword2 -- keyword3
%% https://academic.oup.com/DocumentLibrary/mnras/keywords.pdf

%%%%%%%%%%%%%%%%%%%%%%%%%%%%%%%%%%%%%%%%%%%%%%%%%%

%%%%%%%%%%%%%%%%% BODY OF PAPER %%%%%%%%%%%%%%%%%%

%%%%%%%%%%%%%%%%%%%%%%%%%%%%%%%%%%%%%%%%%%%%%%%%%%%%%%%%%%%%%%%%%%%%%%%%%%%%%%%%%%%%%%%
\section{Introduction}\label{sec:intro} 

%%%%%%%%%%%%%%%%%%%%%%%%%%%%%%%%%%%%%%%%%%%%%
%% nucleosynthesis in 1D CCSNe
%%%%%%%%%%%%%%%%%%%%%%%%%%%%%%%%%%%%%%%%%%%%%
Core-collapse supernova (CCSN) explosion of a massive star chiefly produces elements heavier than oxygen through explosive nucleosynthesis, 
and expel them from the deep inside of the star~\citep[see, e.g.,][and references therein]{2016ApJ...821...38S, 2018ApJS..237...13L}
One of the ultimate goals of CCSN theory is to quantify how heavy elements can be synthesized, how large amounts of these elements are ejected from the star, and 
whether the abundance evolution observed on the surface of Galactic stars is well reproduced with a Galactic chemical evolutions model with the contribution of CCSNe
supplementing with those of other astrophysical sites, such as AGB stars and Type Ia SNe~(see, e.g., \citet{2018MNRAS.476.3432P,2020ApJ...900..179K}).
Addressing these issues requires accurate modeling of neutrino ($\nu$)-radiation-hydrodynamics since the nucleosynthetic yields sensitively depend on not only fluid dynamics but also the $\nu$-radiation field.
However, even the most recent CCSN models suffer from a big uncertainty of collective $\nu$ oscillations, posing a challenge to traditional nucleosynthesis models. 
Offering key insights into the impact of collective $\nu$ oscillations is the subject of this paper.

%%%%%%%%%%%%%%%%%%%%%%%%%%%%%%%%%%%%%%%%%%%%%
%% SFC and FFC and nucleosynthesis
%%%%%%%%%%%%%%%%%%%%%%%%%%%%%%%%%%%%%%%%%%%%%
Detailed theoretical studies of collective $\nu$ oscillations indicate that the slow $\nu$-flavor conversion, one of the collective oscillation modes, would be suppressed deep inside a CCSN core~\citep{2008PhRvD..78h5012E,2011PhRvL.107o1101C,2012PhRvD..85k3007S}.
On the other hand, fast $\nu$-flavor conversion (FFC) potentially overwhelms the matter suppression~\citep[see][for recent reviews]{2020arXiv201101948T,2022arXiv220703561R,2022Univ....8...94C}.
In more recent years, detailed inspections of occurrences of FFCs have been made based on sophisticated multi-dimensional(D) CCSN models, 
and they commonly showed that electron neutrinos lepton number (ELN) crossing, which is a necessary and sufficient condition for the occurrence of FFC, 
appears ubiquitously in the accretion phase of CCSNe ($\lesssim 1 \s$ after the core bounce)~\citep[see, e.g., ][]{2021PhRvD.103f3033A,2021PhRvD.104h3025N,2022ApJ...924..109H}
spurring the interest of FFCs. 
These studies also exhibited that multi-D effects such as asymmetric $\nu$ emission offer favorable circumstances for the occurrence of FFC, 
although it is not clear how much they alter the explosive nucleosynthesis.
The only reference in the literature is \citet{2020ApJ...900..144X}.
The authors conducted nucleosynthetic computations in $\nu$-driven winds at several hundred ms after the bounce, 
incorporating effects of FFCs based on spherically symmetric steady-state wind models.

%%%%%%%%%%%%%%%%%%%%%%%%%%%%%%%%%%%%%%%%%%%%%%%
%%% the present work
%%%%%%%%%%%%%%%%%%%%%%%%%%%%%%%%%%%%%%%%%%%%%%
In this paper, we examine the impact of FFC induced by asymmetric $\nu$ emission on the explosive nucleosynthesis in multi-D CCSN models. 
By employing axisymmetric CCSN models in \citet{2019MNRAS.488L.114F, 2021MNRAS.502.2319F}, we carry out nucleosynthetic computation by incorporating FFC effects. 
The region of FFC is determined by a physically motivated criterion associated with the difference between $\nue$ and $\nueb$ emission.
We adopt a phenomenological flavor-mixing scheme. 
Although the scheme discards non-linear feedback from $\nu$-radiation-hydrodynamics, this study possibly illustrates the essential features of the impact of FFC on nucleosynthesis, 
and will serve as a reference for higher-fidelity CCSN simulations with quantum kinetic $\nu$ transport.

This paper is organized as follows. In section~\ref{sec:SN explosion}, 
we briefly summarize the method and results of axisymmetric CCSN simulations in our previous works~\citep{2019MNRAS.488L.114F, 2021MNRAS.502.2319F}. 
In section~\ref{sec:ffc}, we present our prescription by which to include the effects of FFC on nucleosynthetic computations.
All results are encapsulated in section~\ref{sec:nucleosynthesis} with in-depth analyses of rolls of FFC on explosive nucleosynthesis.
Finally, we discuss the limitation of the present study in section ~\ref{sec:limitation} and 
summarize our conclusion in section~\ref{sec:conclusion}.

%%%%%%%%%%%%%%%%%%%%%%%%%%%%%%%%%%%%%%%%%%%%%%%%%%%%%%%%%%%%%%%%%%%%%%%%%%%%%%%%%%%%%%%
\section{Hydrodynamic models}\label{sec:SN explosion} 

In this study, we do not address all issues relevant to FFCs. 
Rather we focus on the roles of FFC on nucleosynthesis in CCSN. 
To this end, we adopt our CCSN hydrodynamic models~\citep{2019MNRAS.488L.114F}, 
in which the effects of FFCs are neglected. 
We start with briefly summarizing the essence of our CCSN models.

We performed hydrodynamic simulations of CCSN from core collapse to runaway shock expansion for the $19.4\Ms$ progenitor with the solar metallicity~\citep{2002RvMP...74.1015W}.
We employed two codes: GR1D~\citep{2015ApJS..219...24O} in collapsing phase (spherically symmetric simulations) and 
a modified Zeus 2D code~\citep{1992ApJS...80..753S,1992ApJS...80..791S, 2006ApJ...641.1018O, 2007ApJ...667..375O, 2011ApJ...738...61F}
for axisymmetric simulations in the post-bounce phase.
Approximate $\nu$ transport is adopted for the modified Zeus 2D code with light-bulb prescription.

In the post-bounce phase, we excise the central region ($\le 50\km$), where more accurate treatments of neutrino-radiation-hydrodynamic simulations are needed.
The gravity is treated as a central point source, and the mass (representing the mass of the proto-neutron star (NS)) evolves with mass flux through the inner boundary.
In our light-bulb $\nu$-transport, the $\nu$ average energies are assumed to be spherically symmetric, 
which is a reasonable approximation indicated by more elaborate CCSN simulations~\citep{2019ApJ...880L..28N}.
On the other hand, $\nu_e$ and $\bar{\nu}_e$ luminosities, $L_{\nue}$ and $L_{\nueb}$, respectively, can have large dipole components, 
meanwhile the $\nu_x$ and $\nuxb$ luminosities, $L_{\nux}$ and $L_{\nuxb}$, respectively, are assumed to be spherically symmetric. 
In our approach, the degree of dipole component is controlled by $m_{\rm asy}$ as,
\begin{eqnarray}
L_{\nue} &=& L_{\nue, \rm ave} (1 +m_{\rm asy} \cos \theta), \label{eq:asymmetric ne luminosity} \\
L_{\nueb} &=& L_{\nueb, \rm ave} (1 -m_{\rm asy} \cos \theta), \label{eq:asymmetric neb luminosity}
\end{eqnarray}
where $L_{\nue, \rm ave}$ and $L_{\nue, \rm ave}$ denote angular-averaged luminosities.
For the sake of simplicity, the ratios of $\nu$ luminosities and average energies are assumed to be constant among models;
$L_{\nueb, \rm ave}/L_{\nue, \rm ave} = 1$, $L_{\nux}/L_{\nue, \rm ave} = L_{\nuxb}/L_{\nue, \rm ave} = 1/2$, and $\epsilon_{\nueb}/\epsilon_{\nue} = \epsilon_{\nux}/\epsilon_{\nue} = \epsilon_{\nuxb}/\epsilon_{\nue} = 7/6$.
and $\epsilon_{\nue}$, $\epsilon_{\nueb}$, $\epsilon_{\nux}$, and $\epsilon_{\nuxb}$ and the average energy of $\nue$, $\nueb$, $\nux$, and $\nuxb$.
To include the time-dependent features of $\nu$ luminosity and average energy, we use a $\nu$-core model, which is essentially the same prescriptions in \citet{2012ApJ...757...69U}.
The two parameters associated with the $\nu$-core model are tuned so as to reproduce SN1987A-like explosion in the case with $m_{\rm asy} = 0\%$~\citep[see Appendix A of][for more details]{2021MNRAS.502.2319F}.
We ran the simulations for $m_{\rm asy}$ as 0\%, 10/3\%, 10\%, 30\%, and 50\%~\footnote{
Recent $\nu$-radiation-hydrodynamic simulations suggest that asymmetries of $\nu$ emission is $\lesssim 10 \%$~\citep[see, e.g., ][]{2014ApJ...792...96T, 2019ApJ...880L..28N, 2019MNRAS.489.2227V,2021MNRAS.500..696N}.}
It is worth noting that $m_{\rm asy}$ is a pivotal factor to determine the spatial region where FFC occurs, as described in section~\ref{sec:ffc}.

%%%%%%%%%%%%%%%%%%%%%%%%%%%%%%%%%%%%%%%%%%%%%%%%%%%%%%%%%%%%%%%%%%%%%%%%%%%%%%%%%%%%%%%
\section{Prescription to include effects of FFC in nucleosynthetic computations}\label{sec:ffc}

Given the time-dependent fluid background, we carry out post-processing computations of nucleosynthesis with effects of FFCs. 
At present, however, global FFC simulations are unfeasible~\citep[but see][]{2022arXiv220604097N}
and no reliable approximations to include the effects of FFC have been established yet. 
We, hence, employ an ad-hoc approach, which is essentially the same as those used in \citet{2020ApJ...900..144X,2021PhRvL.126y1101L,2022PhRvD.105h3024J,2022PhRvD.106j3003F},
but we alter the FFC criterion to suit CCSNe.

According to recent studies of ELN crossings in CCSN~\citep{2019ApJ...886..139N,2019PhRvD.100d3004A,2021PhRvD.104h3025N}, 
the crossing appears around a $\nu$ sphere when the ratio of number density or flux between $\nu_e$ and $\bar{\nu}_e$ roughly equal each other. 
Our FFC instability criterion is built based on this trend and is given by 
\begin{equation}
0.9 \le \alpha_{\rm f} \le 1.1,
\label{eq:FFC_condition}
\end{equation}
where $\alpha_{\rm f} \equiv F_{\nueb}/F_{\nue}$. $F_{\nue}$ and $F_{\nueb}$ denote the number flux of $\nu_e$ and $\bar{\nu}_e$, respectively.
By using Eqs.~\ref{eq:asymmetric ne luminosity}~and~\ref{eq:asymmetric neb luminosity}, we can express $\alpha_{\rm f}$ as
\begin{equation}
\alpha_{\rm f} = \frac{(L_{\nueb, \rm ave}/\epsilon_{\nueb})}{(L_{\nue, \rm ave}/\epsilon_{\nue})}
\frac{1 -m_{\rm asy} \cos \theta}{1 +m_{\rm asy} \cos \theta} = \frac{6}{7}\frac{1 -m_{\rm asy} \cos \theta}{1 +m_{\rm asy} \cos \theta}, \label{eq:alphafbymth}
\end{equation}
since we set $L_{\nueb, \rm ave} = L_{\nue, \rm ave}$ and $\epsilon_{\nue}/\epsilon_{\nueb} = 6/7$.
Eq.~\ref{eq:alphafbymth} exhibits that the spatial region that satisfies the instability criteria (Eq.~\ref{eq:FFC_condition}) can be determined as functions of $m_{\rm asy}$ and $\theta$.

\begin{figure}
 \begin{center}
  \includegraphics[scale=0.9]{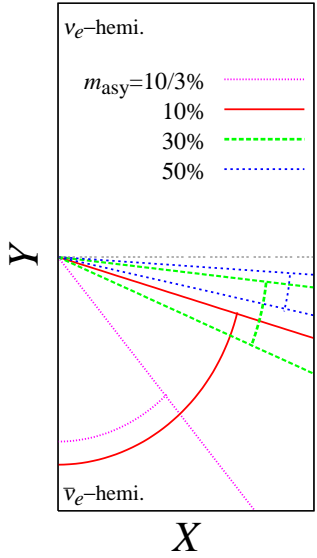}
 \end{center}
 \vspace*{-15pt}
 \caption{
 Spatial regions or angles where the FFC instability criterion (Eq.~\ref{eq:FFC_condition}) is satisfied.
 }
 \label{fig:ffc}
\end{figure}

%%%%%%%%%%%%%%%%%%%%%%%%%%%%%%%%%%%%%%%%%%%%
%%% FFC region
Figure \ref{fig:ffc} portrays the corresponding region where FFC arises. 
In the case with $m_{\rm asy} = 0\%$, no FFCs occur in the entire spatial region, because $\alpha_f$ is 6/7 ($<0.9$). 
In other cases with asymmetric $\nu$ emission, FFC appears in the southern hemisphere, since $\nu_e$ ($\bar{\nu}_e$) emission is weaker (stronger) than that in the northern hemisphere. 
As a result, $\alpha_f$ can be higher than $6/7$ and satisfy the instability criterion ($>0.9$) at a certain angular region. 
It should be mentioned, however, that $\alpha_f$ becomes higher than $1.1$ around the southern pole for cases with $m_{\rm asy} = 30 \%$ and $50 \%$; 
consequently, FFCs are suppressed in these regions. 
As we shall discuss in the next section, the suppression of FFC in these regions is responsible for a non-monotonic dependence on $m_{\rm asy}$ regarding the impact of FFCs on nucleosynthesis.

At present, global FFC simulations are unfeasible~\citep[but see][]{2022arXiv220604097N}
and no reliable approximations to include the effects of FFC have been established yet. 
We, hence, treat the degree of flavor-mixing in a parametric way, which is essentially the same as those used in \citet{2020ApJ...900..144X,2021PhRvL.126y1101L,2022PhRvD.105h3024J,2022PhRvD.106j3003F},
but we alter the FFC criterion to suit CCSNe.
One of the characteristics of FFC is that the flavor conversion is energy-independent\footnote{However, $\nu$-matter interactions may alter the trend; 
\citep[see][for more details]{2022arXiv220709496K}.}.
We, thus, introduce energy-independent parameters, $p$ and $\bar{p}$, which exhibit the survival probabilities of $\nu_e$ and $\bar{\nu}_e$, respectively. 
In terms of $p$ and $\bar{p}$, we express the $\nu$ luminosities after the conversion ($L^{\rm osc}$) as, 
\begin{eqnarray}
& L^{\rm osc}_{\nue} = p L_{\nue} +(1 -p) L_{\nux}, & \label{eq:Lffc_nue} \\
& L^{\rm osc}_{\nueb} = \bar{p} L_{\nueb} + (1 -\bar{p}) L_{\nuxb}, & \label{eq:Lffc_nueb} \\
& L^{\rm osc}_{\nux} = \frac{1}{2} (1-p) L_{\nue} +\frac{1}{2} (1+ p) L_{\nux}, & \label{eq:Lffc_nux} \\
& L^{\rm osc}_{\nuxb}= \frac{1}{2} (1- \bar{p}) L_{\nueb} + \frac{1}{2} (1+ \bar{p}) L_{\nuxb}. & \label{eq:Lffc_nuxb}
\end{eqnarray}
In this study, we adopt five sets of the survival probabilities, $(p, \bar{p})$; 
three symmetric FFC cases, or $(p, \bar{p}) = (1, 1)$ (no FFC), $(2/3, 2/3)$, and $(1/3, 1/3)$ (flavor equilibrium) and two asymmetric FFC cases, $(2/3, 1/3)$ and $(1/3, 2/3)$.
We then carry out the nucleosynthetic computations by the same method as described in our previous papers~\citep{2019MNRAS.488L.114F, 2021MNRAS.502.2319F}, 
but we use $L^{\rm osc}$ in the $\nu$-matter interactions. 
We refer readers to~\citet{2019MNRAS.488L.114F, 2021MNRAS.502.2319F} for the details of our method for nucleosynthetic computation.

%%%%%%%%%%%%%%%%%%%%%%%%%%%%%%%%%%%%%%%%%%%%%%%%%%%%%%%%%%%%%%%%%%%%%%%%%%%%%%%%%%%%%%%
\section{Results}\label{sec:nucleosynthesis}

\begin{figure}
 \includegraphics[scale=0.99]{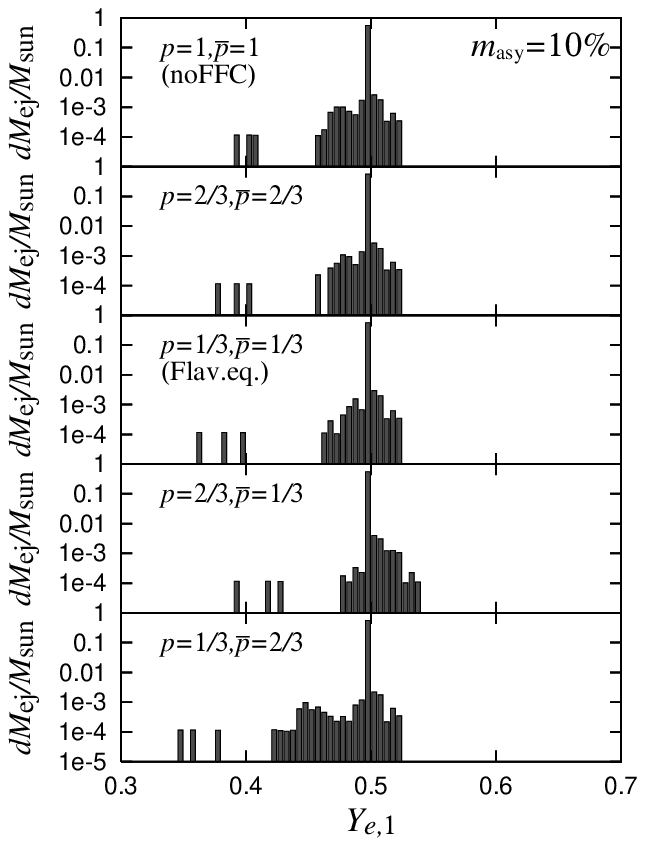}
 \includegraphics[scale=0.99]{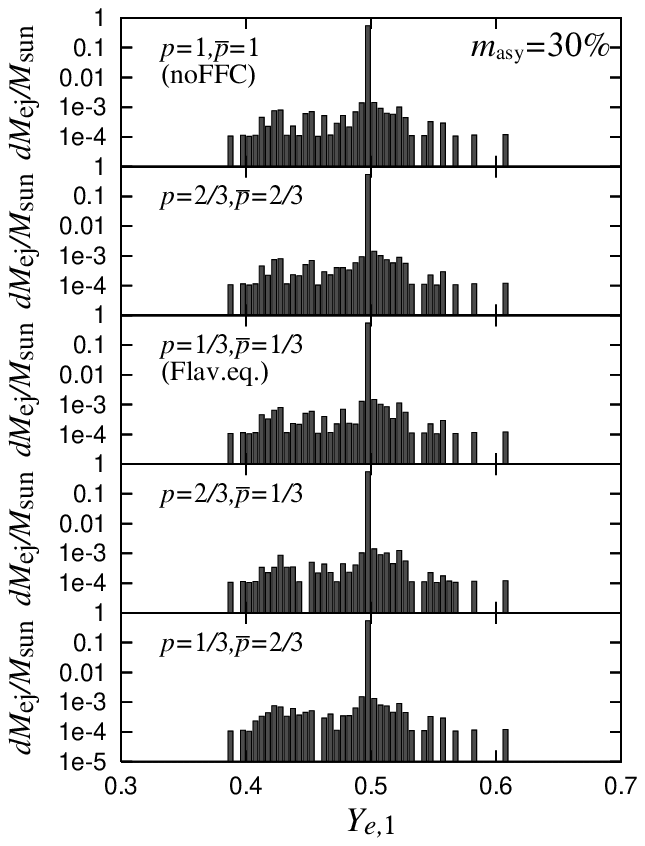}
 \vspace*{-25pt}
 \caption{
 $dM_{\rm ej}$  as a function of freeze-out electron fraction, $Y_{e,1}$.
 We focus on the ejecta that is located at $\le 10,000 \km$ at the onset of gravitational collapse. 
 We show results for $m_{\rm asy}=10 \%$ (top) and $30 \%$ (bottom). 
 In each figure, we display the results with $(p, \bar{p}) = (1, 1)$ (no FFC), $(2/3, 2/3)$, and $(1/3, 1/3)$ (flavor equilibrium), $(2/3, 1/3)$ and $(1/3, 2/3)$ from top to bottom.
 }
 \label{fig:dist-ye}
\end{figure}

%%%%%%%%%%%%%%%%%%%%%%%%%%%%%%%
We present mass distribution in electron fraction, $Y_e$, which is useful to catch the overall trend of the impact of FFC on nucleosynthesis. 
The ejecta mass in a bin of $dY_{e,1} = 0.005$, $dM_{\rm ej}$, is displayed in Figure~\ref{fig:dist-ye}, 
where $Y_{e,1}$ denotes the electron fraction when the temperature of each fluid element becomes $10^9 \rm K$, approximately reflecting the freeze-out value of $Y_e$. 
The dependence of $dM_{\rm ej}$ on $p$ and $\bar{p}$ exhibits the effects of FFCs, and it also varies with $m_{\rm asy}$. 
To see the $m_{\rm asy}$ dependence, we compare the two cases with $m_{\rm asy}=10 \%$ (top) and $30 \%$ (bottom) in Figure~\ref{fig:dist-ye}. 
As shown in the bottom panels, $dM_{\rm ej}$ distribution is less sensitive to $p$ and $\bar{p}$ in the case with $m_{\rm asy}=30 \%$, indicating that effects of FFCs are minor. 
This is due to the fact that FFCs are suppressed around the pole in the southern hemisphere (see Fig.~\ref{fig:ffc}) due to a substantial excess of $\bar{\nu}_e$ compared to $\nu_e$. 
A similar trend is also found in the case with $m_{\rm asy}=50 \%$.

%%%%%%%%%%%%%%%%%%%%%%%%%%%%%%%
In the case of $m_{\rm asy} = 10\%$, the dependence of $dM_{\rm ej}$ distributions on $p$ and $\bar{p}$ is more remarkable than $m_{\rm asy} = 30\%$ (see the top panels in Fig.~\ref{fig:dist-ye}). 
In the symmetric FFC case ($p = \bar{p}$), FFC does not affect $dM_{\rm ej}$ distribution at $Y_{e,1} \gtrsim 0.5$ (see the top three panels). 
This is attributed to the fact that the proton-rich ejecta appears in the $\nu_e$-rich hemisphere, where no FFC arises. 
In the neutron-rich side ($Y_{e,1} \lesssim 0.5$), $dM_{\rm ej}$ distribution tends to shift to lower $Y_{e,1}$ with decreasing $p(=\bar{p})$.
In asymmetric FFC cases ($p \neq \bar{p})$, it substantially deviates from the case with no FFCs.
The ejecta becomes more neutron-rich for $(p, \bar{p}) = (1/3, 2/3)$, while more proton-rich for $(p, \bar{p}) = (2/3, 1/3)$. 
These trends can be understood through the balance of $\nue$ and $\nueb$ absorptions, as we shall discuss the detail with Figure \ref{fig:phys-n-rich-ejecta}. 
It should also be mentioned that the same trend is observed in the case of $m_{\rm asy}=10/3 \%$, albeit less significantly than $m_{\rm asy}=10 \%$. 
This is due to narrower spatial region where FFC occurs (see Fig.~\ref{fig:ffc}). 
%%%%%%%%%%%%%%%%%%%%%%%%%%%%%%%%%%%%%%%%%%%%%%%%%%%%%%%%%%%
%% Ye-Distrubution
%% dYe_ave

\begin{figure}
 \includegraphics[scale=1.15]{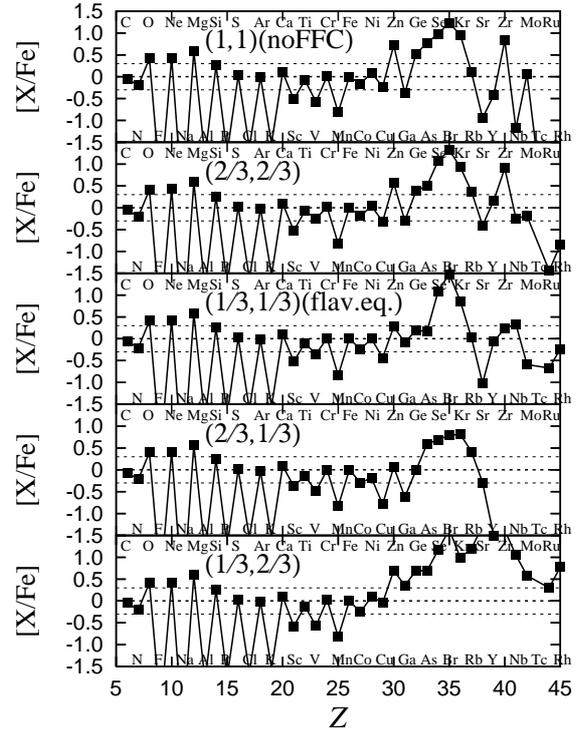}
 \vspace*{-30pt}
 \caption{
 [X/Fe] of all ejecta
 for cases with $m_{\rm asy}=$ 10\% and with $(p, \bar{p}) = (1, 1)$ (no FFC), $(2/3, 2/3)$, and $(1/3, 1/3)$ (flavor equilibrium), $(2/3, 1/3)$ and $(1/3, 2/3)$ in panels from top to bottom.
 }
 \label{fig:xfe}
\end{figure}

%%%%%%%%%%%%%%%%%%%%%%%%%%%%%%%%%%%%%%%%%%%%%%%%%%%%%%%%%%%%%%%%%%%%%%
%%%% [X/Fe] 
Hereafter we focus on the case with $m_{\rm asy}= 10\%$, in which FFC gives the largest impact on nucleosynthesis among our models.
In Figure~\ref{fig:xfe}, the ejecta compositions are displayed as a function of atomic number $Z$, and the abundance pattern is measured with [X/Fe]~\footnote{
$[A/B] \equiv \log \left[ (X_{\rm A}/X_{\rm A, \odot})/(X_{\rm B}/X_{\rm B, \odot})\right]$, where $X_{\rm i}$ and $X_{\rm i,\odot}$ denote a mass fraction of element $\rm i$ and its solar value~\citep{1989GeCoA..53..197A}.}.
We find that FFC facilitates the production of heavier elements ($Z \ge 28$), unless $\bar{\nu}$ flavor conversion is remarkably higher than that of $\nu$ (corresponding to $(p, \bar{p}) = (2/3, 1/3)$ in our model).
This trend is consistent with $dM_{\rm ej}$ distributions in $Y_{e,1}$, i.e., increasing the mass of neutron-rich matter, as displayed in Figure~\ref{fig:dist-ye}. 
On the other hand, elements lighter than Co ($Z \le 27$) other than Sc and V, which are abundantly produced in slightly proton- and neutron-rich ejecta, respectively, are less sensitive to FFCs.
This is attributed to the fact that the mass of the ejecta with $Y_{e,1} \simeq 0.5$, in which the elements lighter than Co are chiefly produced, is less influenced by $\nu$ absorption and thus by FFCs.

\begin{figure}
  \includegraphics[scale=0.85]{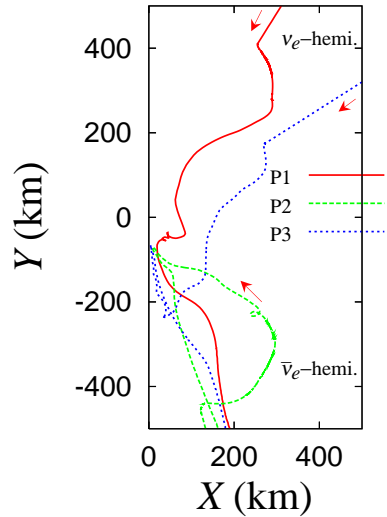}
 \vspace*{-15pt}
 \caption{
 Trajectories of three tracer particles with the lowest $Y_{e,1}$.
 All particles eject through the $\nueb$-hemisphere, where the FFC appears.
 }
 \label{fig:pos-n-rich-ejecta}
\end{figure}

\begin{figure*}
 \includegraphics[scale=1.5]{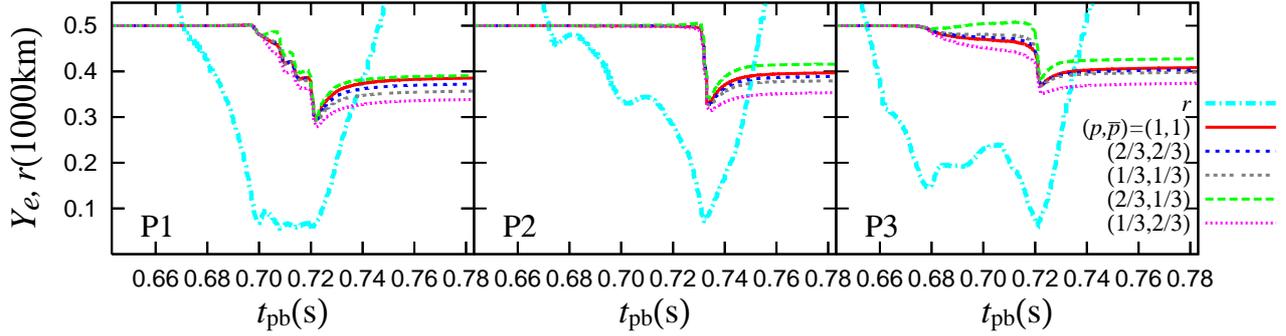} %% P1, P2, and P3
\vspace*{-270pt}
 \caption{Time evolution of $Y_e$ and radial position, $r$, of the three ejecta with the lowest $Y_{e,1}$,
 whose trajectories are shown in Fig. \ref{fig:pos-n-rich-ejecta}, 
 for cases with $m_{\rm asy}=$ 10\% and with $(p, \bar{p}) = (1, 1)$ (no FFC), $(2/3, 2/3)$, $(1/3, 1/3)$ (flavor equilibrium), $(2/3, 1/3)$ and $(1/3, 2/3)$.
 }
 \label{fig:phys-n-rich-ejecta}
\end{figure*}

%%%%%%%%%%%%%%%%%%%%%%%%%%%%%%%%%%%%%%%%%%%
%% Trajectories and Ye/r-evolutions
To understand the mechanism of how FFC gives impact on FFC,
we show the trajectory and the time evolution of $Y_{e}$ for neutron-rich ejecta by focusing on three individual particles having the lowest $Y_{e,1}$. 
In the following discussions, we refer to these particles as P1, P2, and P3 in order of increasing $Y_{e,1}$ (0.390, 0.400, and 0.409 for P1, P2, and P3 in the case without FFC).
Figure \ref{fig:pos-n-rich-ejecta} depicts the trajectories of these particles.
We note that the trajectories are independent of $p$ and $\bar{p}$ for each particle
since the fluid background is identical among models with a different set of $p$ and $\bar{p}$. 
As shown in the figure, all particles reach near the $\nu$ sphere, suggesting that they experience strong deleptonization. 
We also find that they pass through the region around the southern pole, indicating that they are influenced by FFC (see also Fig.~\ref{fig:ffc}). 
%%%%%%%%%%%%%%%%%%%%%%%%%%%%%%%%%%%%%%%%%%%
%% Trajectories

%%%%%%%%%%%%%%%%%%%%%%%%%%%%%%%%%%%%%%%%%%%
%% Ye/r-evolutions
The rolls of FFCs on ejecta compositions can be interpreted through the time evolution of $Y_e$ for these individual particles, which are displayed in Figure~\ref{fig:phys-n-rich-ejecta}. 
As a reference, we also show the radial position of each particle as a blue line. 
Before entering into a detailed discussion of FFC, we briefly describe the essential time-dependent features of $Y_e$ in the case without FFC (see red line in the figure). 
After passing through the accretion shock wave, $Y_e$ of each particle rapidly decreases mainly due to electron capture on protons. 
Once the shock revival is achieved, the ejecta experiences expansion, and then $\nu$ absorption dictates the evolution of $Y_e$. 
In general, $\nu_e$ absorption on neutrons dominates over $\bar{\nu}_e$ one on protons, indicating that $Y_e$ increases with time. 
It should be noted, however, that the increase of $Y_e$ is suppressed in the region with high $\bar{\nu}_e$ emission (southern hemisphere in our models), 
which leads to lower $Y_{e,1}$.
%%%%%%%%%%%%%%%%%%%%%%%%%%%%%%%%%%%%%%%%%%%

Let us now turn our attention to the cases with FFCs. 
As displayed in Figure~\ref{fig:phys-n-rich-ejecta}, the time evolution of $Y_e$ clearly depends on the choice of ($p, \bar{p}$).
The key effect of FFC is the reduction of $\nu$ absorption, which can be understood through the following analysis. 
We start with defining the absorption factor with FFC as, 
$f_{\nue, {\rm osc}} \equiv \epsilon_{\nue}^2 p F_{\nue} +\epsilon_{\nux}^2 (1-p) F_{\nux}$, 
which is roughly proportional to the $\nue$ absorption rate.
We then take the ratio to the case without FFC (i.e., $f_{\nue} = \epsilon_{\nue}^2 F_{\nue}$);
\begin{eqnarray}
 \frac{f_{\nue, {\rm osc}}}{f_{\nue}}
  &=& p(1 +m_{\rm asy} \cos\theta) + \frac{L_{\nux}}{L_{\nue, {\rm ave}}} \frac{\epsilon_{\nux}}{\epsilon_{\nue}} (1-p) \nonumber \\
  &=& \frac{5p +7}{12} +p \, m_{\rm asy} \cos\theta.
  \label{eq:ratio_absorption_factor_nue}
\end{eqnarray}
The first term on the right-hand side of Eq.~\ref{eq:ratio_absorption_factor_nue} is much larger than the second one unless $m_{\rm asy}$ is an order of unity. 
This indicates that the ratio of absorption factor to that without FFC monotonically decreases with $p$, 
exhibiting that the $\nue$ absorption becomes weaker when $p$ is smaller, or the $\nu$-flavor conversion becomes more active. 
The same argument can be applied to $\nueb$. We obtain the following relation,
\begin{eqnarray}
 \frac{f_{\nueb, {\rm osc}}}{f_{\nueb}}
  &=& \bar{p}(1 -m_{\rm asy} \cos\theta) + \frac{L_{\nuxb}}{L_{\nueb, {\rm ave}}} \frac{\epsilon_{\nuxb}}{\epsilon_{\nueb}} (1-\bar{p}) \nonumber \\
  &=& \frac{\bar{p} +1}{2} -\bar{p} \, m_{\rm asy} \cos\theta,
  \label{eq:ratio_absorption_factor_nueb}
\end{eqnarray}
suggesting that $\nueb$ absorption becomes also less efficient for stronger $\nueb$-flavor conversion.

The reduction of $\nu$ absorption leads to two important effects on the evolution of $Y_e$, which depends on $p$ and $\bar{p}$. 
First, the increase of $Y_e$ is suppressed during the ejecta expansion; 
the trend can be seen in the symmetric FFC (see blue and gray lines in Figure~\ref{fig:phys-n-rich-ejecta}). 
This accounts for the increased mass of neutron-rich ejecta. 
Second, the ratio between $\nue$ and $\nueb$ absorptions can be substantially altered in the asymmetric FFCs ($p \ne \bar{p}$). 
More specifically, $Y_e$ tends to be lower in the case of $p<\bar{p}$ (see pink lines in Fig.~\ref{fig:phys-n-rich-ejecta}), 
since $\nue$ absorption is more suppressed than $\nueb$ one. 
In the opposite case $p>\bar{p}$ (see green lines in the same figure), FFC makes the ejecta more proton-rich; 
in fact the maximum $Y_{e,1}$ in the case with $(p,\bar{p})=(2/3,1/3)$ becomes remarkably higher than that without FFC (see Fig.~\ref{fig:dist-ye}).

A few remarks should be made now.
The roles of FFC on nucleosynthesis is qualitatively changed if $L_{\nux} \epsilon_{\nux}$ and $L_{\nuxb} \epsilon_{\nuxb}$ are larger than those of $\nue$ and $\nueb$, respectively.
In these cases, FFC boosts the efficiency of $\nu$ absorption. 
On the other hand, recent CCSN simulations suggest that $L_{\nux} \epsilon_{\nux}$ and $L_{\nuxb} \epsilon_{\nuxb}$ do not overwhelm those of $\nue$ and $\nueb$ during the accretion phase~\citep[see, e.g.,][]{2021MNRAS.500..696N},
suggesting that the present results captures the essential roles of FFC on explosive nucleosynthesis relevant to recent CCSN models. 
Next, \citet{2020ApJ...900..144X} has shown that $\nu$-driven winds become more proton-rich if the effects of FFC are included. 
Their results are consistent with ours, since they adopt $p=0.68$ and $\bar{p}=0.55$ in their simulations, suggesting that $\nueb$ absorption is more suppressed than $\nue$ one by FFCs. 
As such, Eqs.~\ref{eq:ratio_absorption_factor_nue}~and~\ref{eq:ratio_absorption_factor_nueb} are very useful to gaze into the sensitivity of nucleosynthesis on FFC for arbitrary systems.
It should be mentioned, however, that our models neglect the feedback effects of FFC through $\nu$-radiation-hydrodynamics. 
This suggests that more complex effects of FFC on nucleosynthesis may arise in reality. 
In fact, the complex interplay between fluid-dynamics, nucleosynthesis, and FFC has been observed in NS merger systems~\citep{2021PhRvL.126y1101L,2022PhRvD.105h3024J,2022PhRvD.106j3003F}.
We postpone addressing this important issue in future work.

\section{Limitations}\label{sec:limitation}

Though we offer some new insights into the impacts of FFC on explosive nucleosynthesis, the present study has several limitations.
Here, we summarize some methodological shortcomings in the present work.

\subsection{Feedback from FFC to matter evolution}\label{subsec:feedbackFFCtomatter}

In this study, we have performed nucleosynthetic computations in a post-processing manner under given 2D CCSN models in which we neglect the effects of FFCs on fluid dynamics.
It is not self-consistent, and the effects may change the fluid dynamics.  The change might alter the chemical composition of the ejecta.

FFCs influence $\nu$-matter interactions, indicating that they should impact the neutrino heating in the gain region.
As we discussed in section~\ref{sec:nucleosynthesis}, the neutrino absorption is suppressed by FFCs under the current set of parameters for species-dependent neutrino luminosities and average energies.
The suppression implies that the shock revival would be more delayed than our models or could be failed.
If the delay of the shock revival is substantial, the delay results in an appreciable reduction of explosion energy and thus changing nucleosynthesis~\citep{2013ApJ...771...27Y}.

Another thing we do notice as possible feedback from FFC to matter evolution is that FFCs accelerate neutrino diffusion from a proto-NS.
The neutrino diffusion would give positive feedback to the shock revival.
This is because $\nu_e$s are the most abundant flavor of neutrinos inside the proto-NS and convert into $\nu_x$s through FFC,
and then they more easily diffuse out of the proto-NS due to the absence of charged-current reactions for $\nu_x$.
The faster cooling amplified by FFCs has also been observed in radiation-hydrodynamic simulations for the remnant of binary NS merger~\citep{2022PhRvD.105h3024J, 2022PhRvD.106j3003F}.
The higher neutrino diffusion facilitates the proto-NS contraction, which increases the average energy of neutrinos.
Since the neutrino absorption becomes faster with an increasing average energy of neutrinos, the efficiency of neutrino heating would be higher in the gain region~\footnote{The impact is a similar mechanism as the one outlined in \citep[e.g.][]{2018SSRv..214...33B} that many-body effects facilitate shock revival.}, leading to pushing the shock wave further out.

\subsection{2D vs 3D}\label{subsec:2Dvs3D}

There is a consensus that axisymmetric condition in CCSN models exaggerates the anisotropy of matter distributions 
and it may also influence shock revival (see, e.g., \citet{2010ApJ...720..694N, 2012ApJ...755..138H, 2014ApJ...785..123C, 2015ApJ...807L..31L, 2022MNRAS.514.3941N} 
for the argument of the dependence of explodability on spatial dimensions). 
The strong dipolar geometry of ejecta observed commonly in our CCSN models is a representative artifact due to the axisymmetric condition.

Aside from the fluid dynamics, non-axisymmetric components in the neutrino-radiation field may alter the present result. 
As suggested by the recent 3D CCSN simulations, the spatial distribution where ELN crossing occurs has, in general, non-azimuthal structures~\citep[see, e.g., ][]{2021PhRvD.103f3033A,2021PhRvD.104h3025N}, 
and it also tends to spread out more than in 2D. 
These results suggest that the impact of FFCs on nucleosynthesis in 3D would be more complex than what we discuss in the present study. 

It has been shown that the turbulence has no effect on flavor evolution in the accretion phase of CCSN if the neutrino self-interaction is neglected~\citep[see][]{2017arXiv170206951K}, 
On the other hand, ELN crossings are observed at PNS convective layers in recent 3D neutrino-radiation-hydrodynamic simulations~\citep{2020PhRvD.101f3001G}, 
indicating that flavor conversions occur in the strongly turbulent region.
Hence, the non-linear interaction between FFCs and turbulence warrants an in-depth analysis, but we postpone them in future work.

Last but not least, the actual impacts of neutrino quantum kinetics on fluid dynamics and nucleosynthesis can be investigated only by 3D $\nu$-radiation-hydrodynamic simulations 
with the consistent treatment of neutrino transport, flavor conversion (including FFCs), and matter interactions. 
These simulations are common targets for studies of CCSN and the binary NS merger. 
Although remarkable progress on their 3D numerical modeling has been made very recently~\citep{2022arXiv220702214G, 2022PhRvD.105h3024J, 2022PhRvD.106j3003F}, 
there remain many technical issues with them.
We need to keep these uncertainties in mind as caveats in interpreting these results and also ours.

%\vspace*{-12pt}
\section{Conclusion}\label{sec:conclusion}

CCSNe involving asymmetric $\nu$ emission potentially have FFC around a $\nu$ sphere. In this paper, 
we examine the impact of FFC on the explosive nucleosynthesis in such aspherical CCSNe under the assumption that FFC does not affect fluid dynamics.
The main results are summarized as follows.
\begin{enumerate}
 \item FFC appears in the $\nueb$-enhanced region, in which the $\nueb$ luminosity is larger than the $\nue$ one.
       The impact of FFC on nucleosynthesis increases with $m_{\rm asy}$ but it is suppressed for
       large $\nu$ asymmetry ($m_{\rm asy} \gtrsim 30\%$) due to the narrower spatial region where FFC appears.
 \item FFC has little effect on abundances of elements lighter than Co other than Sc and V since the elements are mainly produced in the region where $\nu$-absorption is weak.
 \item For the symmetric FFC cases ($p = \bar{p})$,
       FFC makes the ejecta more neutron-rich. Consequently, the ejecta has larger amounts of elements heavier than Ni (Figs. \ref{fig:dist-ye} and \ref{fig:xfe}). 
       For asymmetric FFC cases ($p \neq \bar{p})$, the ejecta become more neutron- and proton-rich for $p<\bar{p}$ and $p>\bar{p}$, respectively, compared to the case without FFC.
 \item Those dependences of nucleosynthesis on the survival probabilities ($p$ and $\bar{p}$) can be understood through the $\nue$ and $\nueb$ absorption. 
       Eqs.~\ref{eq:ratio_absorption_factor_nue}~and~\ref{eq:ratio_absorption_factor_nueb} provides a simple but essential diagnostics to quantify how FFC changes $\nu$ absorption.
\end{enumerate}

As described in section~\ref{sec:limitation}, there remain crucial improvements in both our CCSN models and nucleosynthetic computations.
These limitations should be kept in mind, especially when considering the application of our findings to astronomical observations. 
Nevertheless, the present study is a necessary step towards connecting collective neutrino oscillations to the theory of CCSN. 
In the future, further understanding of explosive nucleosynthesis will progress according to developments of CCSN simulations with quantum kinetic $\nu$-transport.

\vspace*{-12pt}
\section{Acknowledgements}

We thank the anonymous referee for the valuable comments that helped us to improve our manuscript.
This work is partly supported by JSPS KAKENHI Grant Numbers 20K03957(SF) and 21H01121(SF).
\vspace*{-12pt}

\section*{DATA AVAILABILITY}
The data underlying this article will be shared upon reasonable request to the corresponding author.
\vspace*{-12pt}

%\clearpage

%%%%%%%%%%%%%%%%%%%%%%%%%%%%%%%%%%%

%%%%%%%%%%%%%%%%%%%% REFERENCES %%%%%%%%%%%%%%%%%%

\bibliographystyle{mnras}
\bibliography{ms}

%%%%%%%%%%%%%%%%%%%%%%%%%%%%%%%%%%%%%%%%%%%%%%%%%%

%%%%%%%%%%%%%%%%% APPENDICES %%%%%%%%%%%%%%%%%%%%%

%% \appendix

%% \section{Some extra material}

%% If you want to present additional material which would interrupt the flow of the main paper,
%% it can be placed in an Appendix which appears after the list of references.

%%%%%%%%%%%%%%%%%%%%%%%%%%%%%%%%%%%%%%%%%%%%%%%%%%

%% Don't change these lines
\bsp	% typesetting comment
\label{lastpage}
\end{document}